\input{aipcheck}

\documentclass[
    ,final            
  ]
  {aipproc}

\layoutstyle{8x11double}

\begin{document}

\title{Search for Physics beyond Standard Model at HERA}


\author{Masahiro Kuze for H1 and ZEUS Collaborations}{
  address={Department of Physics, Tokyo Institute of Technology, Tokyo 152-8551 Japan}
}



\begin{abstract}
Recent results on searches for signals of physics beyond Standard Model (SM)
at the $ep$ collider HERA are reviewed.  Limits obtained for contact interaction
models, large extra dimensions and finite quark radius are presented.
Searches for excited-fermion resonances yield unique limits on excited electrons
and neutrinos.  Finally, measurement of $W$ production cross section in $ep$ collision
is presented.
\end{abstract}

\maketitle


\section{Introduction}

HERA was the only $ep$ collider in the world.  It collided 27~GeV electrons or positrons
with 920~GeV protons, yielding a center-of-mass energy of 320~GeV.
Its operation terminated in June last year and each collider experiment, H1 and ZEUS,
collected approximately 0.5~$\rm fb^{-1}$ of data for physics 
analysis.\footnote{The data before the luminosity upgrade in 2000 are referred to as HERA-I
and those afterwards, with longitudinal lepton polarization, are referred to
as HERA-II data.}

\begin{figure}
 \includegraphics[width=\columnwidth]{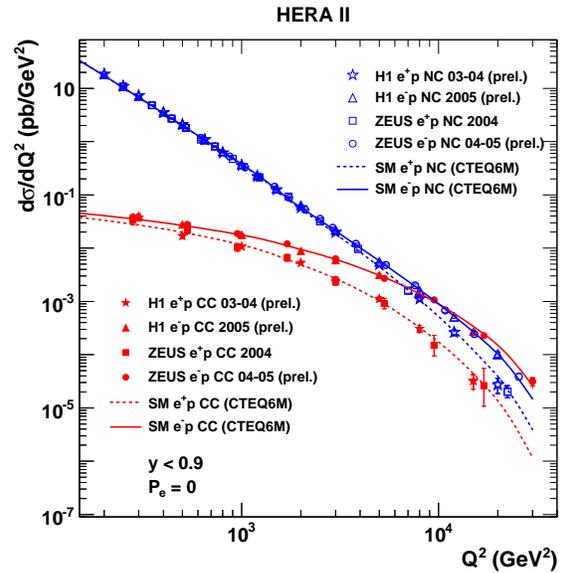}
 \label{NCCC}
 \caption{High-$Q^2$ neutral current (NC) and charged current (CC) cross sections
 measured at HERA.  It is seen that at large values of $Q^2$ the electromagnetic
 and weak interactions are unified and become of the same order of magnitude.}
\end{figure}

At HERA, the lepton and hadron undergo deep inelastic scattering (DIS)
by an exchange of $t$-channel boson ($\gamma$, $Z$ or $W^\pm$),
and at large values of $Q^2$, the momentum transfer squared, it can be
regarded as $eq$ interaction at a very short distance
(see Fig.~\ref{NCCC}).
Hence, by measuring the parton distribution functions using DIS events
at low $Q^2$ and extrapolating them to high $Q^2$ with QCD,
and comparing the SM prediction with the observed events,
one can search for signatures of physics beyond the SM.

In this paper, recent results on searches for contact interactions and
for excited fermions are presented, as well as the most recent result
on $W$ production cross section measurement.
The results on searches for signatures related to supersymmetry
are reported elsewhere in the same conference~{\cite{SS08}.

\section{Contact Interactions and related signatures}
Physics at very high mass scales can appear
at much lower energies as rare processes via virtual effects.
The phenomena can be generally described as contact
interactions (CI), reducing the unknown details of the interaction
to a coupling constant having a dimension of inverse mass.
A classical example is the early days of weak interaction, when
the interaction was considered as a four-fermion CI having a
'small' coupling constant $G_F$ having a dimension of inverse mass
squared.  Later it was clarified that this smallness comes
from the large mass of the propagator boson, the $W$.
At HERA, $eeqq$-CI can affect the neutral current DIS cross sections
at large $Q^2$.  The same CI can also be tested at LEP $e^+e^-$ collider
and TeVatron $p\bar p$ collider.  It is conventional to define the mass
scale $\Lambda$ assuming the value of fundamental coupling constant
at $\sqrt{4\pi}$.

Various models of the new interaction can be considered,
depending on the chiral structure of the coupling to left-
and right-handed leptons and quarks.  Each model
causes different interference with SM interaction,
showing different behavior of NC DIS events at high
$Q^2$ for $e^+p$ and $e^-p$ scattering.

A fit has been made for HERA-II ZEUS data for 
19 CI models, yielding lower limits for the mass
scale $\Lambda$ between 2.0 and 8.0~TeV at 95\% CL.
These results are complementary to those from other colliders
and some are unique limits.
H1 collaboration also derived limits between 1.6 and 5.5~TeV
using HERA-I data.

\begin{figure}
 \includegraphics[width=\columnwidth]{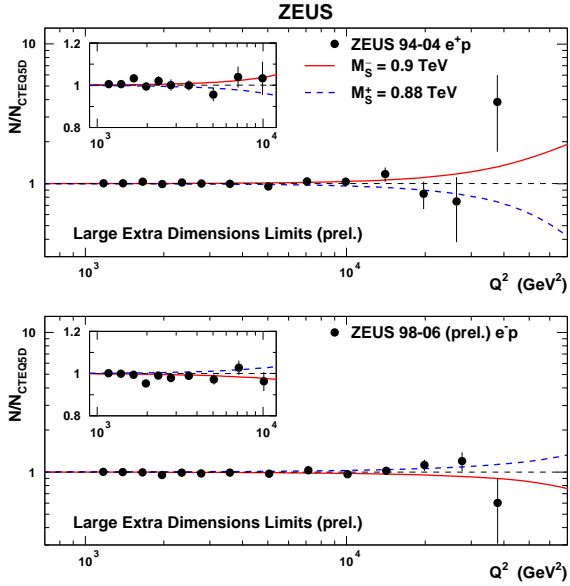}
 \label{led}
 \caption{The points show the ratios of high-$Q^2$ NC DIS events to SM prediction
from ZEUS $e^+p$ (upper) and $e^-p$ (lower) data.  The lines are fits to models
with large extra dimensions.}
\end{figure}
If there are in space-time $n$ extra dimensions which are compactified
to a relatively large scale $R$, and only the gravitational interaction can
propagate in these extra dimensions,  the real GUT scale $M_S$ can
be as low as TeV, much lower than the Planck scale
($R^nM_s^{n+2}\approx M_{\rm Planck}^2$)~\cite{ADD}.
This is an attractive hypothesis to solve the hierarchy problem,
and its collider consequence is that exchange of Kaluza-Klein
excitations of gravitons modifies the SM-particle scattering at high energy.
Phenomenologically, it can be reduced to a CI-like formalism with
$\lambda /M_s^4$ as a parameter.
Fits to HERA-II ZEUS data (HERA-I H1 data), shown in Fig.~\ref{led}, yield lower limits on
$M_s$ of 0.88 (0.82)~TeV for $\lambda$=+1 and 0.90 (0.78)~TeV for $\lambda$=$-$1,
respectively.

Finally, from the same data set, a classical approach on measuring the form factor
of a quark can be performed, repeating the Hofstadter measurement on the
proton but at a scale of $Q^2\approx 40,000~\rm GeV^2$ instead of $1~\rm GeV^2$.
At this scale, the spatial resolution reaches the order of $10^{-16}$~cm, one thousandth of the
proton radius.
If the quark is not elementary but has a finite radius, the scattering cross section
would decrease as the virtual-boson probe 'penetrates' into it, since it begins
to see less electroweak charge.
The upper limit from ZEUS (H1) NC data on the quark size, assuming the electron
is point-like, is $0.62\times10^{-16}$ $(0.74\times10^{-16})$~cm.

\section{Excited Fermions}
It can be seen at each scale of Nature that whenever there is compositeness
(structure) of matter, there are excited states and emission (radiation) phenomena
between the ground and excited states.
For example, excited molecule does light emission (eV scale), excited atoms
emit X-ray (keV), excited nuclei and gamma ray (MeV), and excited nucleons
(resonances) decay with pion (0.1~GeV) emission.
Therefore, if the leptons/quarks are not elementary but composite, their excited
states could radiate gauge bosons (0.1~TeV) when decaying to the ordinary state.

\begin{figure}
 \includegraphics[width=\columnwidth]{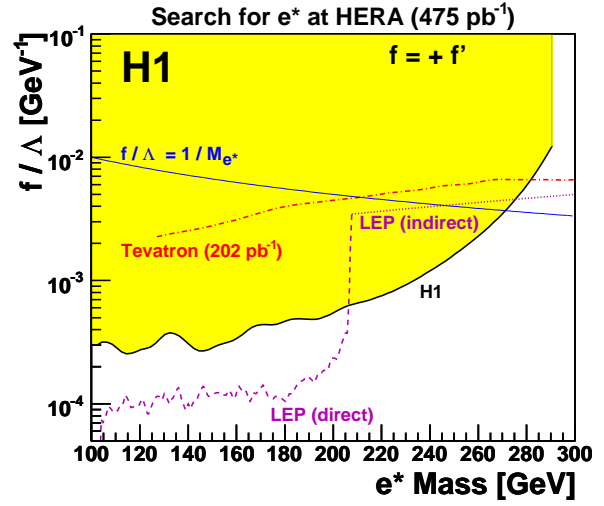}
 \label{estar}
 \caption{Limits on $f/\Lambda$ as a function of excited electron mass obtained from H1 HERA-II
 data.}
\end{figure}
\begin{figure}
 \includegraphics[width=\columnwidth]{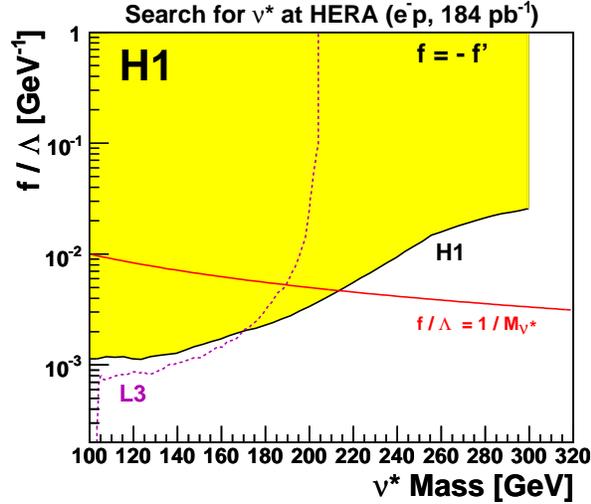}
 \label{nustar}
 \caption{Limits on $f/\Lambda$ as a function of excited neutrino mass obtained from H1 HERA-II
data.}
\end{figure}
Both H1 and ZEUS collaborations search for mass resonances in fermion+boson pairs.
Figure ~\ref{estar} shows the latest results from H1~\cite{H1estar} for an excited electron ($e^*$) search.
The resonance was searched for in $e\gamma$, $\nu W$ and $eZ$ decay modes,
followed by hadronic and leptonic decays of $W/Z$.
The limits are expressed in coupling over composite scale, $f/\Lambda,$ as a
function of the $e^*$ mass $(M_{e^*})$.
Assuming the relation $f/\Lambda = 1/M_{e^*}$, lower limit of $M_{e^*}>272~\rm GeV$
can be obtained, which is more stringent than LEP and TeVatron limits.
The indirect limits from LEP above 200~GeV come from $e^+e^-\to\gamma\gamma$
process.

Figure~\ref{nustar} shows the H1 results~\cite{H1nustar} for an excited neutrino ($\nu^*$) search.
Since the production ($ep\to \nu^* X$) proceeds through $W$ exchange (charged
current), $e^-p$ collision gives much higher sensitivity than $e^+p$.
The search was done for $\nu\gamma$, $eW$ and $\nu Z$ decay modes.
It can be seen that HERA has unique sensitivity for excited neutrinos
in the mass range above 200~GeV.  In the result shown in the figure,
the lower limit on $\nu^*$ mass for the same assumption as above is obtained
to be 216~GeV.

\section{W production}
Finally, the most recent results from ZEUS are presented concerning events
with a topology with a lepton with high transverse momentum ($P_T$) and
missing $P_T$~\cite{ZEUSW}.  Largest SM contribution to this topology comes from on-shell
$W$ production, and excess of such events with additional large hadronic
$P_T (P_T^X)$ could be a signature of e.g. FCNC single-top production with
subsequent decay $t \to bW$.

From the whole HERA data set (504 pb$^{-1}$), ZEUS observed 11 isolated
electron or muon events with $P_T^X$>25~GeV, compared with SM expectation
of 12.9$\pm$1.7 events (of which 77\% comes from $W$ production).  Also in
the lower-$P_T^X$ region the observation was in agreement with SM prediction.
Figure~\ref{W} shows the transverse-mass distribution for electron events, which
shows a clear Jacobian peak expected for $W$ production.
\begin{figure}
 \includegraphics[width=\columnwidth]{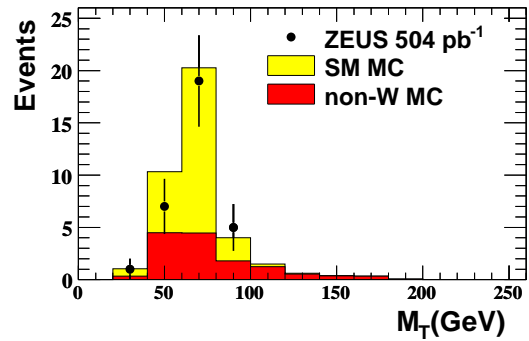}
 \label{W}
 \caption{Distribution of transverse mass for the ZEUS events with high-$P_T$
 isolated electron and missing $P_T$.}
\end{figure}

From these events, the total cross section for $W$ production in $ep$ collisions
was obtained to be 0.89$^{+0.25}_{-0.22}\rm(stat.)\pm$0.10(syst.)~pb.
This is the smallest total cross section measured at HERA, and in agreement
with theoretical prediction at NLO calculation.

\section{Summary}
HERA has ceased data-taking after about 15 years, with $\approx0.5 \rm fb^{-1}$
per experiment of high-energy $ep$ collision data.
The data give solid confidence of perturbative QCD and indispensable
inputs to LHC physics.
The short distance $eq$ interaction has an unique opportunity to search for particles
and forces beyond SM.
Limits on contact interactions, large extra dimensions and quark radius, and
on excited electrons and neutrinos are presented, for example.
Many results are competitive/complementary with other colliders.
The search list is not yet exhaustive, and many results from whole
HERA data are expected to come.
Also a combined-results working group from H1 and ZEUS has been formed.

\begin{theacknowledgments}
 The author is supported by grant-in-aid (Kakenhi) from Japanese MEXT (16001002).
 
\end{theacknowledgments}





\IfFileExists{\jobname.bbl}{}
 {\typeout{}
  \typeout{******************************************}
  \typeout{** Please run "bibtex \jobname" to optain}
  \typeout{** the bibliography and then re-run LaTeX}
  \typeout{** twice to fix the references!}
  \typeout{******************************************}
  \typeout{}
 }

\end{document}